\documentclass[twocolumn]{revtex4-2}
\usepackage{graphicx}
\usepackage{dcolumn}
\usepackage{bm}
\usepackage{amsmath}
\usepackage{amsfonts}
\usepackage{amssymb}
\usepackage{cancel}
\usepackage{color}
\usepackage{epstopdf}
\usepackage{float}
\usepackage{placeins}
\providecommand{\U}[1]{\protect\rule{.1in}{.1in}}

\begin{document}
\title{A diagrammatic representation of entropy production}
\author{Andr\'es Vallejo}
\affiliation{\begin{small} Facultad de Ingenier\'{\i}a, Universidad 
de la Rep\'ublica, Montevideo, Uruguay\end{small}}
\date{\today}
\begin{abstract}
\noindent We introduce a natural way of visualizing the entropy production in heat
transfer processes between a simple thermodynamic system and a thermal reservoir. 
This representation is particularly useful to highlight the asymmetric 
character of the heating and cooling processes, when they are analyzed from the 
second-law point of view. 

\end{abstract}

\maketitle

\section{Introduction}

Among the different theories that study the human cognitive structure, 
there is a consensus that the use of multiple representations (such as 
analogies, diagrams, graphs, formulas, simulations, etc.) facilitates 
learning processes \cite{Treagust}. In physics, empirical evidence shows 
that students that are able to effectively combine different kinds of 
representations usually develop better problem-solving skills and, consequently, a 
more profound conceptual understanding \cite{,Munfaridah, Dufresne,Kohl,De Cock}. 

In particular, diagrammatic representations, when well designed, have been recognized 
as a valuable tool for understanding and analyzing physical phenomena. 
They complement the analytical results, revealing in some cases aspects of the 
phenomena studied that do not follow directly from the equations. For example, 
in classical mechanics, the use of energy bars is an effective tool to visualize energy 
transformations and conservation \cite{van Heuvelen}. In special relativity, 
Minkowski diagrams provide a natural explanation of counter-intuitive phenomena such as 
the length contraction \cite{Mermin}. Of course, the list is much longer.

In the study of thermodynamics, pressure-volume and temperature-entropy diagrams, 
among others, are traditionally employed to represent 
the states and processes undergone by thermodynamic systems. Although their understanding by 
students is not exempt from difficulties \cite{Pollock}, its employment highlights the 
relationships between the thermodynamic properties of the system and its energetic interactions 
with the environment, heat and work, which, in some cases can be represented as areas in those diagrams.

Following this philosophy, in this paper we introduce a natural way of 
visualizing the entropy production in heat transfer processes between a simple 
thermodynamic system and a thermal reservoir. According 
to the second law of thermodynamics, the global entropy variation associated with a physical 
process is always non-negative, and it only vanishes in the 
(idealized) case in which the process is reversible. This implies that, unlike energy, which only flows 
from one system to another, entropy is produced in any real physical process. Operationally, since the total amount of generated entropy (usually called \textit{entropy production}) coincides with the global entropy variation, it can be obtained by partitioning the universe in a convenient manner and adding the corresponding entropy changes. These quantities can be positive, negative or zero, but their sum is always non-negative. Since it only vanishes for reversible processes, entropy production can  
be considered a measure of departure of the process from the reversible limit, which, in turn, is 
directly linked to the lost work that the system could have performed under 
reversible conditions. For this reason, it is a crucial 
quantity in thermodynamic analysis.

In recent decades notable advances have been obtained regarding the explanation 
of the observed behavior of entropy in macroscopic systems, and the conditions 
under which violations of entropy increase principle can occur have been established \cite{Evans,Croocs,Jarzynski,Seifert}.
However, and despite the difficulties that entropy analysis presents for students, alternative representations 
of entropy production are absent in classic textbooks \cite{vanWylen,Cengel,Zemansky,Callen,Moran}, and a diagrammatic visualization of this quantity is, to the best of our knowledge, still lacking (with the remarkable exception of \textit{stochastic line integrals}, which characterize irreversibility in noisy driven non equilibrium systems and can be related to entropy production \cite{Ghanta}).   

The representation proposed is simple, since it is based on the fundamental 
theorem of calculus, and, in particular, on the possibility of finding the variation of 
a function in an interval as the integral of its derivative in the same interval. 
However, unlike physical quantities that are trivially defined as integrals, such as impulse 
or work, representing the entropy production in this way is, as we shall see, a slightly more 
subtle task.

The outline of this work is as follows. In Section II, we obtain a generic expression 
for the entropy production during a thermalization process with a heat reservoir. In 
Section III we show, through the analysis of three concrete examples, how from the previous 
result we can generate diagrams in which the entropy production can be 
naturally represented as the area of a certain region. 
Finally, some remarks and conclusions are presented in Section IV.

\section{Entropy production in a thermalization process}

Let us consider the thermalization process undergone by a simple thermodynamic 
system ($A$) that exchanges energy with a thermal reservoir ($B$). We will focus 
on the particular set of processes in which some thermodynamic property of system $A$, 
that we denote $Y$, adopts the same value in the initial and final states, i.e.:
	\begin{equation}\label{Y1=Y2}
	Y_2=Y_1.
	\end{equation}
Note that $Y$ can be the temperature, the pressure, or any other property 
satisfying Eq. (\ref{Y1=Y2}) (the limitations imposed by the adoption of this 
hypothesis will be discussed further below). If we consider another independent 
property $X$ of system $A$, the state postulate for a single-phase simple 
compressible substance ensures that the rest 
of the properties of the system are 
functions of $X$ and $Y$. In particular, for the entropy and the internal 
energy, we can write:
	\begin{equation}
	S^{\text{A}}=S^{\text{A}}(X,Y); \hspace{0.2cm}U^{\text{A}}=U^{\text{A}}(X,Y).
	\end{equation}
In what follows we will assume that the behaviour of the substance is simple 
enough for the previous functions to admit a mathematical description. However, 
the results obtained will be equally valid regardless of whether such a model 
exists.

Now let us focus on the determination of the degree of irreversibility of the 
thermalization process computing the 
global entropy production. In order to obtain the entropy variation of $A$ we 
will make use of the 
fact that entropy is a state function, so its variation is independent on the 
process followed by the system between the same initial and final states.
This allows us to consider blue an internally reversible trajectory
linking those states and 
find the entropy variation integrating 
the differential $dS$ along that trajectory. The simplest 
trajectory satisfying the condition (\ref{Y1=Y2}) is 
the curve $Y=Y_1=const.$, so we can write the entropy variation of $A$ as:	

	\begin{equation}\label{delta S^A}
	\Delta S^{\text{A}}_{1\rightarrow 2}=\int_{X_1}^{X_2}\dfrac{\partial S^{\text{A}}}{\partial X}\biggr{\vert}_{Y=Y_1}dX
	\end{equation}
Notice that it is not necessary that the property $Y$ is constant during the 
real process: in fact, in many cases of interest $Y$ is not even well defined 
during the whole process. 
	
Assuming that the environment $B$ is an internally reversible heat reservoir, 
its entropy variation can be written as:

	\begin{equation}\label{delta S^B def}
	\Delta S^{\text{B}}_{1\rightarrow 2}=\dfrac{Q^{\text{B}}}{T^{B}},
	\end{equation}
where the heat exchange between the systems can be obtained from the first law 
applied to system $A$:

	\begin{equation}\label{Q^A}
	Q^{\text{B}}=-Q^{\text{A}}=-W^{\text{A}}-\Delta U^{\text{A}}_{1\rightarrow 2},
	\end{equation}	 	
where $W^{\text{A}}$ is the work performed by system $A$, and $\Delta U^{\text{A}}_{1\rightarrow 2}$ 
can be found analogously to $\Delta S^{\text{A}}_{1\rightarrow 2}$:
	
	\begin{equation}\label{delta U^A}
	\Delta U^{\text{A}}_{1\rightarrow 2}=\int_{X_1}^{X_2}\dfrac{\partial U^{\text{A}}}{\partial X}\biggr{\vert}_{Y=Y_1}dX
	\end{equation}
Combining Eqs.  (\ref{delta S^B def}), (\ref{Q^A}) and (\ref{delta U^A}), we obtain:
	\begin{equation}\label{delta S^B}
	\Delta S^{\text{B}}_{1\rightarrow 2}=-\dfrac{1}{T^{B}}\left[\int_{X_1}^{X_2}\dfrac{\partial U^{\text{A}}}{\partial X}\biggr{\vert}_{Y=Y_1}dX+W^A\right].
	\end{equation}
Finally, from Eqs. (\ref{delta S^A}) and (\ref{delta S^B}), we can write the following general formula for the global entropy variation (note the simplification in the notation of the constant property in the partial derivative):
	\begin{equation}\label{delta S^{Univ}}
\Delta S^{\text{Univ}}_{1\rightarrow 2}=\int_{X_1}^{X_2}\left[\dfrac{\partial S^{\text{A}}}{\partial X}\biggr{\vert}_{Y}-\dfrac{1}{T^{B}}\dfrac{\partial U^{\text{A}}}{\partial X}\biggr{\vert}_{Y}\right]dX-\dfrac{W^{\text{A}}}{T^{B}}.
	\end{equation}

Instead of proceeding automatically and calculate all the above quantities in order to 
obtain $\Delta S^{\text{Univ}}_{1\rightarrow 2}$, in the next section we will 
show how the presence of integrals in the above expression can be exploited to develop 
a useful geometric interpretation of the global entropy production. Of course, the 
examples that we will analyze can be studied without the need to apply Eq. 
(\ref{delta S^{Univ}}), simply carrying out the corresponding entropy analysis.

\section{Applications}		

\subsection{Case 1: $V_1=V_2$}
\subsubsection*{Example: Thermalization of an incompressible solid}
\begin{figure}[!h]
	{\includegraphics[scale=0.5,angle=0, clip]{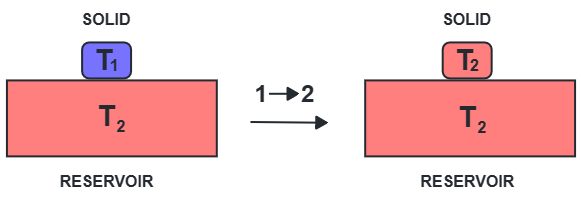}}
	\caption{Thermalization of an incompressible solid due to heat exchange with a thermal reservoir}
	\label{Fig1}
	\end{figure}

Let us consider 1 kilogram of an incompressible solid (system $A$) with constant 
heat capacity $C$. The solid, initially at temperature $T_1$, is placed in contact 
with a thermal reservoir at temperature $T_2>T_1$ (see Fig. (\ref{Fig1})). 
Since the solid is incompressible, 
the volume is constant, so it adopts the same value in the initial and the final states 
and, consequently, it can adopt the role of the property $Y$. On the other hand, it can be seen that, 
for an incompressible solid, the internal energy and the pressure are basically functions of the temperature, 
so we choose the temperature $T$ as the property $X$. 

With these choices, the partial derivative of the internal energy appearing in Eq. 
(\ref{delta S^{Univ}}) is nothing but the heat capacity of the system:
	\begin{equation}\label{dU/dX_case1}
	\dfrac{\partial U^{\text{A}}}{\partial T}\biggr{\vert}_{V}=C,
	\end{equation}
which, by hypothesis, is constant. Consequently, the internal energy satisfies that
	\begin{equation}
	dU^{\text{A}}=CdT,
	\end{equation}
and from the Gibbs relation:
	\begin{equation}
	TdS=dU+PdV,
	\end{equation}
and the incompressibility condition, we obtain that
	\begin{equation}\label{dS/dX_case1}
	\dfrac{\partial S^{\text{A}}}{\partial T}\biggr{\vert}_{V}=\dfrac{C}{T}.
	\end{equation}
	
\begin{figure}[!ht]
 {\includegraphics[scale=0.55, clip]{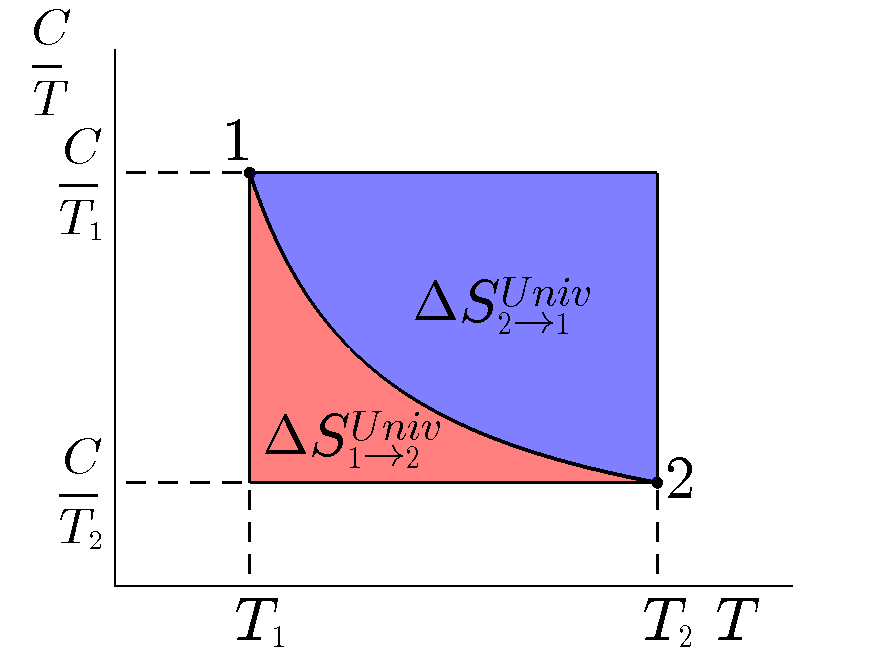}}
 \caption{Entropy production associated to the cooling (blue, upper region) and heating (red, lower region) processes undergone by an incompressible solid with constant heat capacity. Both processes are implemented by
placing the solid in contact with a thermal reservoir. From the diagram we conclude that the cooling process is always more irreversible, for all possible values of the temperatures involved.}
 \label{Fig2}
\end{figure}

Noting that the work performed is zero, from Eqs. (\ref{delta S^{Univ}}), 
(\ref{dU/dX_case1}) and (\ref{dS/dX_case1}), we obtain that the entropy 
produced until the equilibrium is reached can be written as
	\begin{equation}
	\Delta S^{\text{Univ}}_{1\rightarrow 2}=\int_{T_1}^{T_2}\dfrac{C}{T}dT-\dfrac{1}{T_2}\int_{T_1}^{T_2}CdT.
	\end{equation}
For reasons of convenience that will become clear to the reader shortly, we decided to 
explicitly perform only the second integral, after which we finally obtain that:

	\begin{equation}\label{DS12_case1}
	\Delta S_{1\rightarrow 2}^{\text{Univ}}=\int_{T_1}^{T_2}\dfrac{C}{T}dT-\dfrac{C}{T_2}(T_2-T_1)
	\end{equation}
Now let us focus on the graphic interpretation of the above equation. First, note that the 
first term represents the area under the graphic of the function 
	\begin{equation}
	f(T)=\dfrac{\partial S^{\text{A}}}{\partial T}\biggr{\vert}_{V}=\dfrac{C}{T}
	\end{equation}
in the interval $[T_1,T_2]$ (see Fig. (\ref{Fig2}). On the other hand, the quantity $(C/T_2)(T_2-T_1)$ can be 
interpreted as the area of a rectangle of height $C/T_2$, and whose base is the mentioned interval. Consequently, 
since the global entropy production is the difference between those quantities, it is 
represented by the red (lower) area indicated in Fig. (\ref{Fig2}). 

What happens if we place the system in thermal contact with a reservoir at the original 
temperature $T_1$? Applying Eq. (\ref{delta S^{Univ}}) for the cooling process from 
temperature $T_2$ to temperature $T_1$, we obtain that the entropy produced is
	\begin{equation}
	\Delta S^{\text{Univ}}_{2\rightarrow 1}=\int_{T_2}^{T_1}\dfrac{C}{T}dT-\dfrac{1}{T_1}\int_{T_2}^{T_1}CdT,
	\end{equation}
which, after performing the second integral, adopts the form:

\begin{equation}\label{DS21_case 1}
	\Delta S^{Univ}_{2\rightarrow 1}=\dfrac{C}{T_1}\left(T_2-T_1\right)-\int_{T_1}^{T_2}\dfrac{C}{T}dT.
	\end{equation}
In this case the entropy production is the difference between the area of the 
rectangle of height $C/T_1$ and the area under the graphic of $f(T)$, so it 
corresponds to the blue (upper) area in Fig. (\ref{Fig2}). 

A clear virtue of this representation is that it evidences the asymmetry between 
the heating and cooling processes from the point of view 
of the irreversibility involved. The analysis of Fig. \ref{Fig2} shows that, due to the 
convex character of $f(T)$, cooling a solid by energy exchange with a reservoir is always 
a more irreversible process that heating by the same 
procedure. This asymmetry is particularly remarkable for large temperature differences, 
since while Eq. (\ref{DS12_case1}) behaves like
	\begin{equation}
	\Delta S_{1\rightarrow 2}^{\text{Univ}}\simeq C\log{\dfrac{T_2}{T_1}},
	\end{equation}
showing a logarithmic behaviour, the cooling process is approximately 

	\begin{equation}
	\Delta S_{2\rightarrow 1}^{\text{Univ}}\simeq C\dfrac{T_2}{T_1},
	\end{equation}
so its growth is linear with $T_2/T_1$. Graphically, this can be appreciated by noting 
the different rate at which both areas grow as $T_1$ approaches zero or $T_2$ tends to 
infinity.

\FloatBarrier
\subsubsection*{Generalization: Isochoric process}
The above result can be generalized for an arbitrary substance that undergoes 
an isochoric process and reaches thermal equilibrium with a reservoir at temperature 
$T_2$. First, we must note that the relation 
	\begin{equation}\label{C_V/T}
	\dfrac{\partial S^{\text{A}}}{\partial T}\biggr{\vert}_{V}=\dfrac{C_V}{T}
	\end{equation}
is a thermodynamic identity valid for any substance \cite{Cengel,vanWylen}. Taking into 
account the definition of 
heat capacity at constant volume:
	\begin{equation}\label{C_V}
	C_V=\dfrac{\partial U^{\text{A}}}{\partial T}\biggr{\vert}_{V},
	\end{equation}
and noting that for this kind of processes the work performed is zero, from Eqs. 
(\ref{delta S^{Univ}}), (\ref{C_V/T}) and (\ref{C_V}), we conclude that
	\begin{equation}
	\Delta S^{\text{Univ}}_{1\rightarrow 2}=\int_{T_1}^{T_2}\left[\dfrac{C_V}{T}-\dfrac{C_V}{T_2}\right]dT.
	\end{equation}
The above equation shows that the entropy production in any constant volume process 
corresponds to the area between the curves $C_V/T$ and $C_V/T_2$, when they are plotted 
as a function of $T$ (since in the general case $C_V$ is a function of $T$ and $V$, it is necessary to substitute the correct volume in order to plot the functions). 

\begin{figure}[!h]
 {\includegraphics[scale=0.55, clip]{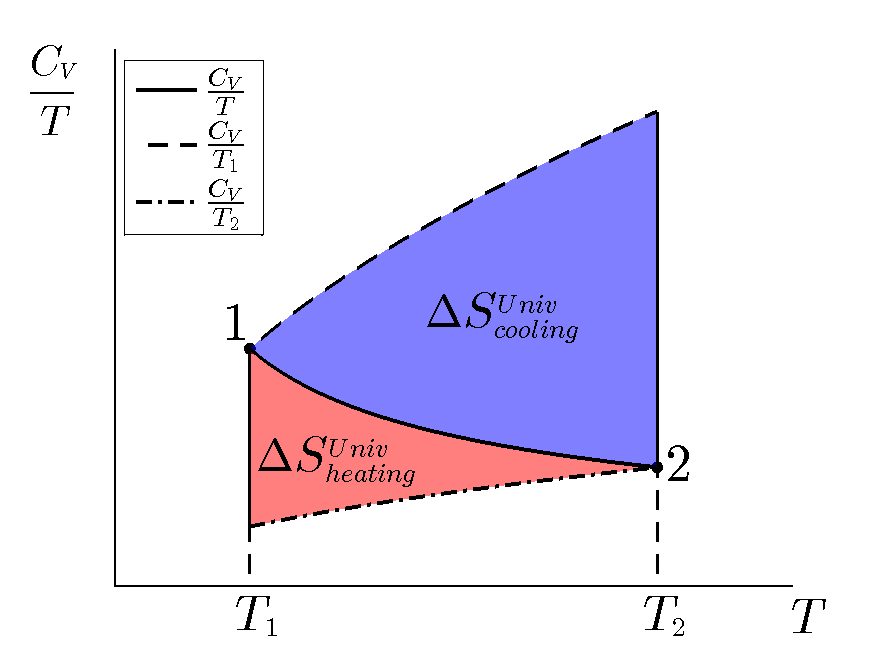}}
 \caption{Entropy production diagram for the isochoric cooling (blue, upper region) and heating (red, lower region) processes of a substance that exchanges heat with thermal reservoirs. This particular diagram corresponds to a substance such that $C_V$ is proportional to $\sqrt{T}$ in the considered region.}
 \label{Fig3}
\end{figure}

It is possible to verify that if the system is cooled to 
the initial temperature $T_1$ employing a thermal reservoir, the total entropy increase is:
\begin{equation}
	\Delta S^{\text{Univ}}_{2\rightarrow 1}=\int_{T_1}^{T_2}\left[\dfrac{C_V}{T_1}-\dfrac{C_V}{T}\right]dT,
	\end{equation}
so graphically it corresponds to the area between the functions $C_V/T_1$ and $C_V/T$. Both quantities are represented in Fig. (\ref{Fig3}) for a particular dependence of $C_V$ with the temperature.

\subsection{Case 2: $P_1=P_2$}
\subsubsection*{Example 1: Quasi-static expansion of a perfect gas}	

\begin{figure}[!h]
	{\includegraphics[scale=0.5,angle=0, clip]{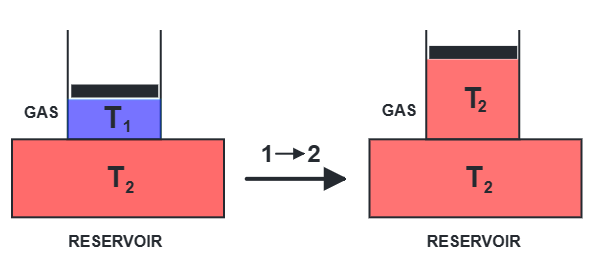}}
	\caption{Quasi-static expansion of a perfect gas due to heat exchange with a thermal reservoir}
	\label{Fig4}
	\end{figure}

\begin{figure}[!h]
	{\includegraphics[scale=0.55, clip]{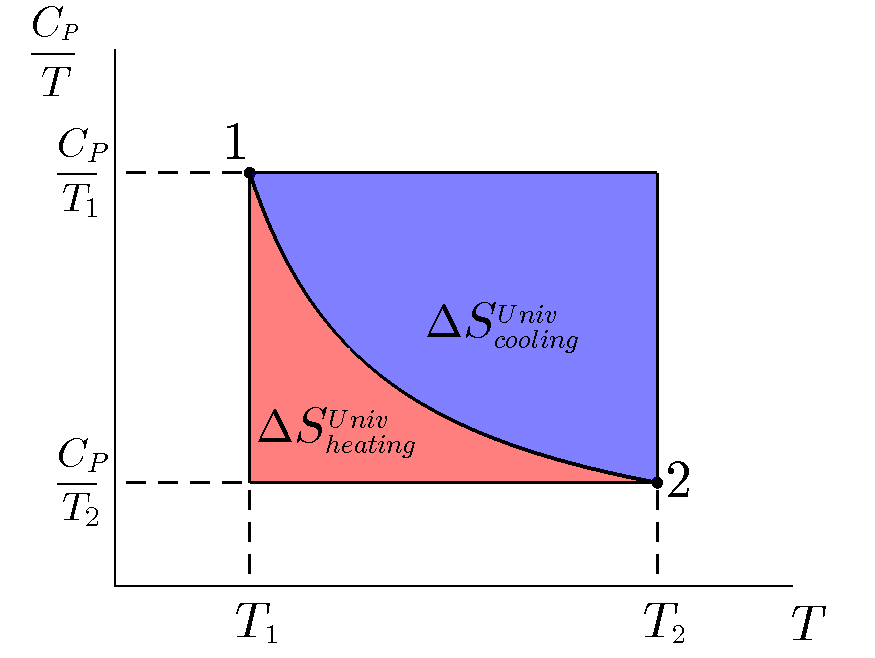}}
	\caption{Entropy production diagram for the isobaric (quasi-static) compression and expansion of a perfect gas. The diagram is analogous to Fig. (\ref{Fig2}), but with $C_P$ playing the role of $C_V$.}
	\label{Fig5}
	\end{figure}

As a second example, let us consider 1 kilogram of a perfect gas ($C_P,C_V$ const.) 
contained in a piston-cylinder device. The gas is initially in an equilibrium state 
at temperature $T_1$ and pressure $P_1$, product of the combined effect of the 
atmosphere and the weight of the piston. In what follows we analyze the process 
undergone by the gas when it is placed in contact with a thermal reservoir at 
temperature $T_2>T_1$ (see Fig. (\ref{Fig4}). We will suppose that the rate of energy transfer between 
the gas and the reservoir is slow enough so that it is reasonable to assume that 
the expansion of the gas occurs at constant pressure. 
Since the internal energy of an ideal gas depends only on the temperature, and taking 
into account that the pressure of the gas is constant, it is clear that the temperature 
and the pressure can play the role of the properties $X$ and $Y$. From the relation 
$dU=C_VdT$, we obtain:
	\begin{equation}\label{dU/dX_case2}
	\dfrac{\partial U^{\text{A}}}{\partial T}\biggr{\vert}_{P}=C_V.
	\end{equation}
Consequently, from the Gibbs relation
	\begin{equation}\label{Gibbs}
	TdS=dH-VdP,
	\end{equation}
and using that $dH=C_PdT$, we obtain that
	\begin{equation}\label{dS/dX_case2}
	\dfrac{\partial S^{\text{A}}}{\partial T}\biggr{\vert}_{P}=\dfrac{C_P}{T}.
	\end{equation}
On the other hand, the ideal gas satisfies the equation
	\begin{equation}\label{ideal}
	PV=mRT,
	\end{equation}
where $m$ is the mass, and R is the universal gas constant (divided by the molar mass).
Using this result, we can write the work performed by 1 kilogram of gas as:
\begin{equation}\label{W_case2}
W^{\text{A}}=P_1(V_2-V_1)=R(T_2-T1).
\end{equation}
Then, from Eqs. (\ref{delta S^{Univ}}), (\ref{dU/dX_case2}), 
(\ref{dS/dX_case2}) and (\ref{W_case2}), we obtain that
	\begin{equation}
	\Delta S^{\text{Univ}}_{1\rightarrow 2}=\int_{T_1}^{T_2}\dfrac{C_P}{T}dT-\dfrac{1}{T_2}\int_{T_1}^{T_2}C_VdT-\dfrac{R(T_2-T_1)}{T_2}.
	\end{equation}
Finally, performing the second integral and using that, for an ideal gas, the relation 
$R+C_{V}=C_P$ holds, we conclude that
	\begin{equation}
	\Delta S^{\text{Univ}}_{1\rightarrow 2}=\int_{T_1}^{T_2}\dfrac{C_P}{T}dT-\dfrac{C_P(T_2-T_1)}{T_2},
	\end{equation}
This result is analogous to the one obtained in the previous subsection, so the entropies generated during the processes of heating and cooling of the gas can be identified as areas, but in this case in the $C_P/T-T$ diagram. The asymmetry in irreversibility between both processes can also be appreciated (see Fig. \ref{Fig5}).  

\FloatBarrier
\subsubsection*{Example 2: Irreversibilities in the Brayton cycle}
The Brayton cycle is a thermodynamic cycle used in gas turbine engines and some air-breathing 
jet engines. In aeronautical applications, the implementation of the Brayton cycle is not strictly a cycle, 
since the working fluid (in this case, air) is taken in from the surroundings and, after transiting several stages, is discharged to the atmosphere. Here we focus in the study of the closed Brayton cycle, which operates in a closed loop by recirculating the gas after it is cooled down.

The cycle is composed of four processes: adiabatic compression, isobaric heating, adiabatic 
expansion, and isobaric cooling (see Fig. (\ref{Fig6})). The high-pressure gas coming from the compressor (state 1) is directed into a combustion chamber, where it is mixed with fuel and ignited at constant pressure. For simplicity, we model this stage as an isobaric heat exchange with a thermal reservoir. Next, the high enthalpy gas (state 2) expands adiabatically in the turbine, producing power. Finally, the exit gas from the turbine (state 3) is cooled and send to the compressor (state 4). We assume that the compressor and the turbine are isentropic, so the only irreversibilities in the cycle are associated with the heat transfer between the gas (assumed perfect) and the reservoirs. In what follows we will evaluate these irreversibilities by computing the entropy productions (per unit of mass flowing) in the heater and the cooler.

It is important to emphasize that, when analyzing a particular control volume (instead of the universe), the entropy variation rate and the entropy production rate are no longer the same. While the former can be both positive and negative (or zero), the latter is always non-negative, and it is the one associated with the irreversibility produced in that control volume. Of course, since when considering the universe as a whole, the entropy variation and entropy generation rates coincide, once a partition is defined, we can obtain the global entropy production both by adding the entropy variation rates of the parts, or by adding the corresponding entropy generation rates.

Once the cycle reaches the stationary regime, the energy and the entropy in any element of the cycle are constant in time ($\frac{dE}{dt}=\frac{dS}{dt}=0$), so applying the first and the second law as a rate equation for a control volume enclosing the heater, we obtain:
	\begin{figure}[!h]
	{\includegraphics[scale=0.9, clip]{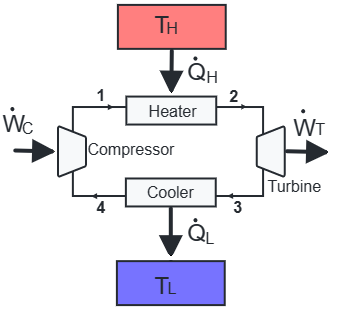}}
	\caption{Schematic representation of a closed Brayton cycle.}
	\label{Fig6}
	\end{figure}	
	\begin{equation}
	\cancelto{0}{\dfrac{dE^{Heater}}{dt}}=\dot{m}(h_1-h_2)+\dot{Q}_{H}-\cancelto{0}{\dot{W}_{H}}
	\end{equation}
and
	\begin{equation}
	\cancelto{0}{\dfrac{dS^{Heater}}{dt}}=\dot{m}(s_1-s_2)+\dfrac{\dot{Q}_{H}}{T_H}+\dot{S}_{gen}^{Heater},
	\end{equation}
where the kinetic and potential energy variations were discarded. Combining the above equations with Eq. (\ref{Gibbs}), we obtain that the entropy production in the heater (per unit of mass flowing) is:
	\begin{equation}
	s^{Heater}_{gen}=\dfrac{\dot{S}^{Heater}_{gen}}{\dot{m}}=\int_{T_1}^{T_2}\dfrac{C_P}{T}dT-\dfrac{C_P(T_2-T_1)}{T_H},
	\end{equation}
where $T_H$ is the temperature of the hot reservoir. Following the same procedure, for the cooling 
process we obtain:
	\begin{equation}
	s^{Cooler}_{gen}=\dfrac{C_P(T_3-T_4)}{T_L}-\int_{T_4}^{T_3}\dfrac{C_P}{T}dT,
	\end{equation}
where $T_L$ is the temperature of the cold reservoir. The graphic representation of these quantities as areas is similar to the previous cases and can be seen in Fig. (\ref{Fig7}). Note that, for the construction of the diagram, it is only required to know the operating temperatures and the value of $C_P$.

	\begin{figure}[!h]
	{\includegraphics[scale=0.6, clip]{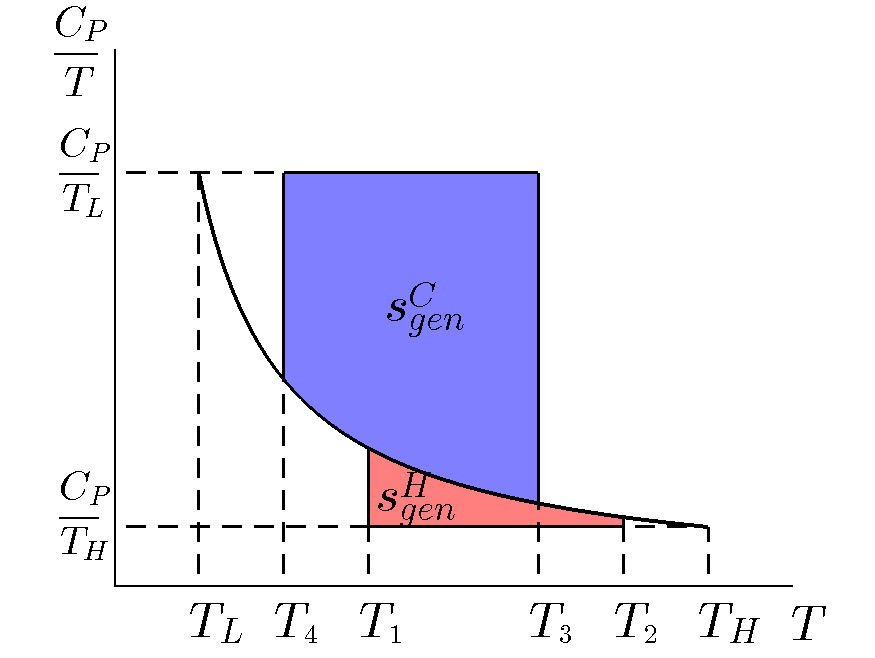}}
	\caption{Diagrammatic representation of the entropy production (per unit of mass flowing) in the heater (red, upper region), and in the cooler (blue, lower region) for a Brayton cycle. The global entropy production is represented by the sum of both areas.}
	\label{Fig7}
	\end{figure}

\subsubsection{Generalization: Isobaric process}
The generalization of the above results to any isobaric process is similar to 
the derivation presented for the isochoric process. Noting that the relation 
	\begin{equation}\label{C_P/T}
	\dfrac{\partial S^{A}}{\partial T}\biggr{\vert}_{P}=\dfrac{C_P}{T}
	\end{equation}
is valid for any substance, and that the first law for an isobaric process reads 
	\begin{equation}
	Q^A=\Delta H^{A},
	\end{equation}
it is possible to show that the global entropy production due to the thermalization process 
with a reservoir at temperature $T_2$ is: 
	\begin{equation}
	\Delta S^{\text{Univ}}_{1\rightarrow 2}=\int_{T_1}^{T_2}\left[\dfrac{C_P}{T}-\dfrac{C_P}{T_2}\right]dT ,
	\end{equation}
so the entropy production can be naturally interpreted as the area between the curves $C_P/T$ and $C_P/T_2$ (again, since $C_P$ may be a function if $P$ and $T$, it could be necessary to replace the value of $P$ in order to plot the functions).	The details of this derivation are left to the reader.


\subsection{Case 3: $T_1=T_2$}
\subsubsection{Abrupt compression and expansion of a perfect gas}
The previous examples seem to suggest that the proposed method can only 
be applied if the property $Y$ is effectively constant during the process. 
However, this is not the case: the method works as long as the property $Y$ 
takes the same value at the beginning and at the end of the process (in fact, 
it could even not be well defined during a part of the process).

\begin{figure}[!h]
	{\includegraphics[scale=0.5, clip]{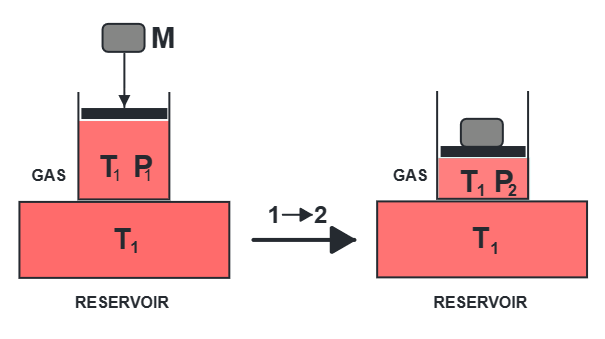}}
	\caption{Abrupt compression of a perfect gas, The mass $M$ added to the 
	piston is such that $P_2A=Mg+P_0A$, where $A$ is the section of the piston, 
	and $P_0$ is the atmospheric pressure. This implies that, in the long term, 
	the gas reaches an equilibrium state at pressure $P_2$. }
	\label{Fig8}
	\end{figure}

To provide an example, let us consider 1 kilogram of a perfect gas in a 
piston-cylinder device at pressure $P_1$, in equilibrium with a thermal 
bath at temperature $T_1$. The gas is subjected to an abrupt compression 
by suddenly adding a weight on the piston. The weight is such that the action 
of the atmospheric pressure and the piston would be equilibrated by an internal 
pressure of value $P_2>P_1$ (see Fig. (\ref{Fig8})). The cylinder is kept in 
contact with the bath 
during the process, so, after transient oscillations, the gas will reach a new 
equilibrium state at pressure $P_2$ and at the initial temperature, due to 
heat exchange with the bath \cite{Mungan2017}. 

\begin{figure}[!h]
	{\includegraphics[scale=0.55, clip]{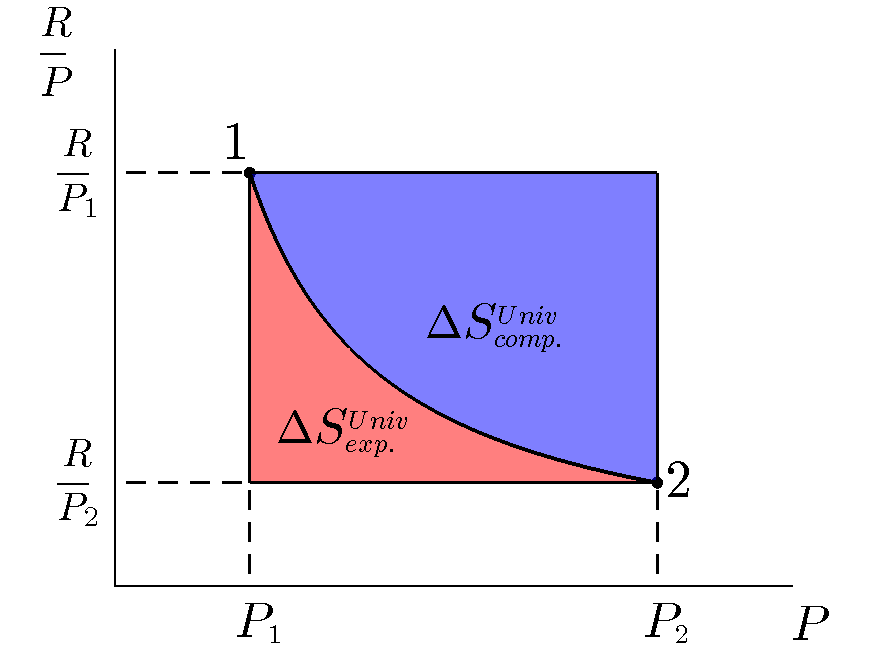}}
	\caption{Area representation of the entropy productions corresponding to the abrupt compression and expansion processes of a perfect gas.}
	\label{Fig9}
	\end{figure}

According to this reasoning, a natural option is to choose the pressure $P$ 
as the property $X$, and the temperature $T$ as $Y$. 
Since $U$ depends only on the temperature ($dU^{\text{A}}=C_VdT$), 
we have that
	\begin{equation}\label{dU/dX_case3}
	\dfrac{\partial U^{\text{A}}}{\partial P}\biggr{\vert}_{T}=0,
	\end{equation}
and from the Gibbs relation, Eq. (\ref{Gibbs}), combined with the equation of 
state, Eq. (\ref{ideal}), we conclude that
	\begin{equation}\label{dS/dX_case3}
	\dfrac{\partial S^{\text{A}}}{\partial P}\biggr{\vert}_{T}=-\dfrac{R}{P}.
	\end{equation}
On the other hand, the gas is compressed by a constant external pressure of value 
$P_2$, so the work performed by the gas is \cite{Gislason}:
	\begin{equation}\label{W_case3}
	W^{A}=P_2(V_2-V_1)=RT_1\left(1-\dfrac{P_2}{P_1}\right),
	\end{equation}
where the equation of state was again employed. The global entropy production can be found 
by combining Eqs. (\ref{delta S^{Univ}}), (\ref{dU/dX_case3}), (\ref{dS/dX_case3}) and 
(\ref{W_case3}), the result is:
	\begin{equation}\label{delta S^univ_case3}
	\Delta S^{\text{Univ}}_{1\rightarrow 2}=\dfrac{R(P_2-P_1)}{P_1}-\int_{P_1}^{P_2}\dfrac{R}{P}dP.
	\end{equation}
From a similar reasoning, the reader may verify that by removing the weight from the piston and 
waiting for the system to reach the new equilibrium state at the initial pressure, the entropy 
increases in an amount
	\begin{equation}
	\Delta S^{\text{Univ}}_{2\rightarrow 1}=\int_{P_1}^{P_2}\dfrac{R}{P}dP-\dfrac{R(P_2-P_1)}{P_2},
	\end{equation}
As expected, both quantities represent areas, but in this case on the diagram  $R/P-P$ (see Fig. (\ref{Fig9})). As a consequence, the abrupt compression in contact with a reservoir is a much 
more irreversible process than the abrupt expansion under the same conditions. 

Finally, note that in this case is not possible to derive a generalization, 
since different works can be performed during an isothermal process. However, from the knowledge 
of the substance and of the particular process followed by the system, it would certainly be 
possible to link the entropy generated with the area of a region in an appropriate diagram.

\section{Remarks and conclusions}

The diagrammatic representation of entropy production presented in this work 
provides a clear and intuitive way of visualizing the irreversibility 
involved in heat transfer processes. The fact that the global entropy production 
is obtained as the difference between two areas evidences that systems in 
thermal interaction have entropy variations of opposite signs, but such that 
their addition is always positive. The examples provided demonstrate the 
usefulness of this method, and particularly its ability 
to highlight the non-symmetric character of the heating and cooling processes, 
when they are analyzed from the second-law perspective. 

As mentioned, the method assumes that there exists a property ($Y$)
that adopts the same value in the initial and final states. This condition is 
necessary so that the entropy and the internal energy of the system can be 
expressed as integrals in a single variable. However, this assumption is not a 
serious restriction, since, combining the usual properties, it is always possible 
to define a new property satisfying this condition. Of course, this procedure may 
generate a more complex dependence of the entropy and internal energy on
the property $X$ selected.

It is important to note that, as a general rule, the curves appearing in entropy 
production diagrams 
introduced in this paper are not representations of the actual processes undergone 
by the system. In fact, in some cases the independent property $X$ might 
not even be well-defined during some stages of the process (for example, this occurs 
with the pressure in the abrupt compression/expansion). As explained in \cite{Anacleto}, 
a process is representable as a curve in a diagram if and only if it is quasi-static, 
and an abrupt process does not satisfy this condition. On the contrary, 
the curves shown in the entropy production diagrams are, in each case, representations of 
the internally reversible process at constant $Y$ that links the same initial and final
states that the real process. This hypothetical process may, of course, but does not have to, 
coincide with the real process.

We insist that the use of Eq. (\ref{delta S^{Univ}}) throughout the article is due to reasons 
of systematization and space; in practice, it might be more instructive to carry 
out the corresponding entropy analysis each time, and then interpret the terms as 
areas in the appropriate diagram.

Since the knowledge of calculus in two variables is the only mathematical prerequisite 
to apply this method (in fact, in one 
variable it would be sufficient), we believe that this approach is appropriate to 
be included in the teaching of thermodynamics at the undergraduate level. 
Likewise, since it provides a complementary perspective to the calculations, we believe 
that this tool could be valuable for all those who carry out thermodynamic analysis in 
different contexts.

\section*{Acknowledgments}
This work was partially supported by Agencia Nacional de Investigación e Innovación and Programa de Desarrollo de las Ciencias Básicas (Uruguay).

\section*{Author Declarations}
The author has no conflicts to disclose.

\end{document}